\documentclass[pra,twocolumn,showpacs,superscriptaddress]{revtex4}
\usepackage[latin1]{inputenc}
\usepackage{graphicx}
\usepackage{amssymb}
\usepackage{amsmath}
\usepackage{xspace}
\usepackage{bm}
\usepackage{color}
\usepackage{ulem}

\newfont{\tensy}{cmsy10}

\newcommand{\ie}[0]{i.e.\@\xspace}

\newcommand{\nag}{{\phantom{\dag}}}

\newcommand{\las}[0]{\langle}
\newcommand{\ras}[0]{\rangle}

\begin{document}

%%%%%%%%%%%%%%%%%%%%%%%%%%%%%%%%%%%%%%%%%%%%%%%%%%%%%%%%%%%%%%%%%%%%%%%%%%%%%
 %%%%%%%%%%%%%%%%%%%%%       TITLE & ABSTRACT          %%%%%%%%%%%%%%%%%%%%%%%
%%%%%%%%%%%%%%%%%%%%%%%%%%%%%%%%%%%%%%%%%%%%%%%%%%%%%%%%%%%%%%%%%%%%%%%%%%%%%

\title{Polariton Mott insulator with trapped ions or circuit QED}

\date{\today}

\author{M. Hohenadler}
\affiliation{Institut f\"ur Theoretische Physik und Astrophysik, Universit\"at W\"urzburg, 97074 W\"urzburg, GER}

\author{M. Aichhorn}
\affiliation{Institut f\"ur Theoretische Physik -- Computational Physics, TU Graz, 8010 Graz, AUT}

\author{L. Pollet}
\affiliation{Institut f\"ur Theoretische Physik, ETH Zurich, 8093 Zurich, SUI}

\author{S. Schmidt}
\affiliation{Institut f\"ur Theoretische Physik, ETH Zurich, 8093 Zurich, SUI}

\begin{abstract} 
  We consider variants of the Jaynes-Cummings-Hubbard model of lattice
  polaritons, taking into account next-nearest-neighbor, diagonal and
  long-range photon hopping in one and two dimensions. These models are
  relevant for potential experimental realizations of polariton Mott
  insulators based on trapped ions or microwave stripline resonators.  We
  obtain the Mott-superfluid phase boundary and calculate excitation spectra
  in the Mott phase using numerical and analytical methods. Including the
  additional hopping terms leads to a larger Mott phase in the case of
  trapped ions, and to a smaller Mott phase in the case of stripline
  resonators, compared to the original model with nearest-neighbor
  hopping only. The critical hopping for the transition changes by up to
  about 50 percent in one dimension, and by up to about 20 percent in two
  dimensions. In contrast, the excitation spectra remain largely unaffected.
\end{abstract} 


\maketitle

\section{Introduction}\label{sec:intro}

The study of strongly correlated and condensed phases of ultra-cold atoms has
become an extremely active and versatile research field over the last decade.
It has been triggered by the experimental realization of a Mott insulator to
superfluid quantum phase transition of neutral atoms in an optical lattice
\cite{Gr.Ma.Es.Ha.Bl.02} as described by the seminal Bose-Hubbard model
\cite{PhysRevB.40.546}. The recent realization of strong light-matter
interaction in various cavity-quantum electrodynamics (QED) systems has
triggered an immense interest in realizing strongly correlated and condensed
phases with photons as well.  Bose-Einstein condensation and superfluidity of
weakly interacting polaritons, \ie, quasiparticles that form when photons
strongly interact with matter, have already been observed experimentally with
exciton-polaritons in a quantum well coupled to a photonic crystal cavity
\cite{Polbec}.

Today, a key challenge is to reach the limit of strong correlations, that is
a photonic or polaritonic Mott insulator \cite{GrTaCoHo06,HaBrPl06,AnSaBo07}.
The Jaynes-Cummings-Hubbard model (JCHM) has been introduced to describe such
an extreme state of light in an array of coupled high-$Q$ electromagnetic
resonators \cite{GrTaCoHo06}. In the generic JCHM, photons hop between
nearest-neighbor (nn) cavities and interact with a single two-level system
locally in each cavity. This fundamental light-matter coupling introduces a
non-linearity into the system, leading to an effective repulsive
photon-photon interaction. The competition between hopping (delocalization)
and light-matter interaction (localization) leads to a phase diagram
featuring Mott lobes similar to the Bose-Hubbard model.  The phase diagram,
excitations and critical exponents of the generic JCHM have been studied
theoretically using various numerical and analytical methods
\cite{GrTaCoHo06,HaBrPl06,AnSaBo07,Ro.Fa.07,Ai.Ho.Ta.Li.08,Zh.Sa.Ue.08,Ko.LH.09,Sc.Bl.09,Sc.Bl.10,PhysRevB.81.024301,PhysRevB.82.045126,HoAiScPo11}
and it has been shown that the same critical theory applies to the JCHM and
Bose-Hubbard model \cite{Ko.LH.09,Sc.Bl.09,Sc.Bl.10,HoAiScPo11}.  Original proposals for an
experimental realization of the JCHM were based on nitrogen vacancy centers
in diamond \cite{GrTaCoHo06} or self-assembled quantum dots in photonic
crystals \cite{Na.Ut.Ti.Ya.07}. However, both variants suffer from short
coherence times and are prone to disorder effects.

Two rather recent proposals, which are not limited with respect to
decoherence and disorder, are based on circuit QED
\cite{Ko.LH.09,NuKo11,WuGa11} and trapped ions \cite{Ivanov1,Ivanov2}. In
circuit QED, Josephson qubits are coupled to stripline resonators. Arrays can
be formed by capacitively coupling resonators in one-dimensional (1D) or
two-dimensional (2D) geometries. In trapped ions, the role of photons is
played by phonon excitations which are exchanged between the ions, and arrays
can be formed either in linear Paul traps or in microtraps.  However, in both
cases, photon transfer beyond the canonical nn hopping cannot be neglected.
In a circuit QED setup, the 2D square lattice geometry gives rise to
next-nearest neighbor (nnn) hopping.  In the case of trapped ions, hopping is
mediated by the Coulomb interaction between dipoles, and is thus long ranged
in all lattice dimensions.  An accurate quantitative description of the
Mott-superfluid transition in these systems requires more general versions of
the JCHM. In this paper, we show that additional hopping terms substantially
modify the Mott-superfluid phase boundary.  We compare exact numerical
results obtained from extensive quantum Monte Carlo (QMC) simulations with
the variational cluster approach (VCA) and analytical calculations based on a
linked-cluster expansion.

The paper is organized as follows. In Sec.~\ref{sec:model}, we introduce the
JHCM with a general hopping term, and specify the cases pertaining to circuit
QED arrays and trapped ions. In Sec.~\ref{sec:methods}, we outline the
numerical and analytical methods used. Section~\ref{sec:results} contains a
discussion of our results for the effect of additional hopping terms on the
phase diagram and the elementary excitations. A summary is given in
Sec.~\ref{sec:summary}. Finally, the Appendix contains relevant
results for the atomic limit.

\section{Models}\label{sec:model}

\begin{figure}[t]
  \includegraphics[width=0.45\textwidth,clip]{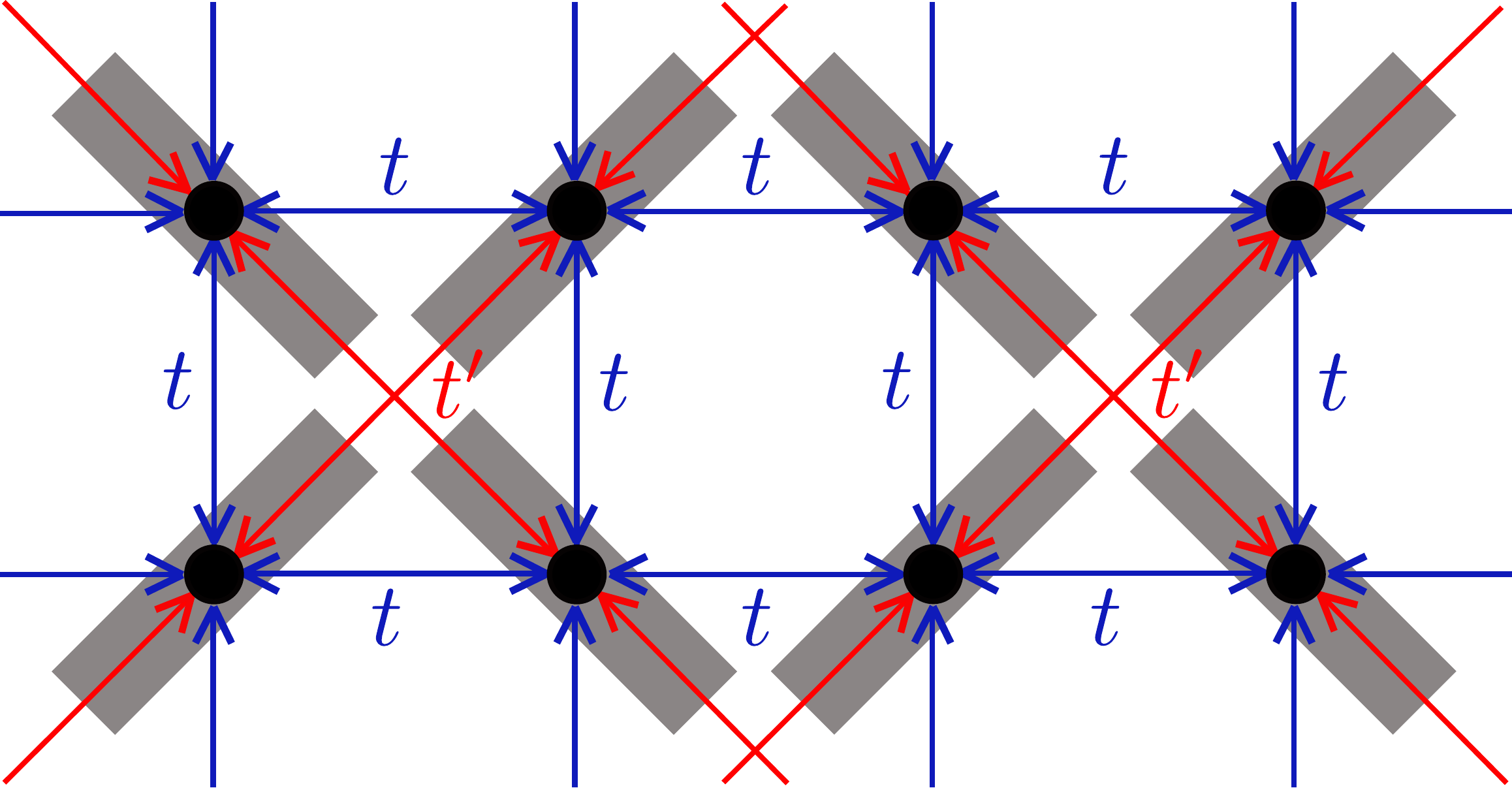}
  \caption{\label{fig:lattice}
    (Color online)
    Circuit-QED realization of the JCHM on a 2D square lattice. A filled circle indicates
    a lattice site, defined by the position of a qubit. Each qubit is located inside
    a resonator represented by a rectangle. The capacitive coupling between
    resonators gives rise to photon hopping across the lattice. Two different
    couplings occur: nearest-neighbor hopping with amplitude $t$ along the
    lattice axes, and diagonal next-nearest neighbor hopping {\it in every other}
    plaquette with amplitude $t'$.
  }
\end{figure}

The JCHM is defined by the Hamiltonian
\begin{equation}\label{eq:ham}
  \hat{H}
 =  \sum_i h_i^{\rm JC} 
   +\sum_{ij} t_{ij}(a^\dag_i a^\nag_j+\text{H.c.}) 
   - \mu \hat{N}
  \,,
\end{equation}
with the on-site JC Hamiltonian for the cavity at site $i$ of the lattice,
\begin{equation}\label{eq:ham3}
 h_i^{\rm JC}  = \omega_\text{b} a^\dag_i a^\nag_i
  +
  \omega_\text{q} \sigma^+_i \sigma^-_i
  +
  g (\sigma^+_i a^\nag_i + \sigma^-_i a^\dag_i)
  \,.
\end{equation}
Here, $\omega_\text{b}$ denotes the energy of bosons (photons or phonons),
and $\omega_\text{q}$ is the level splitting of the qubits (ions or Josephson
qubits).  The spin operators $\sigma^\pm$ describe intrasite transitions
between the two qubit levels induced by emission or absorption of a boson
with rate $g$. The bosonic operators $a,a^\dag$ fulfill the usual commutation
relations $[a_i,a^\dag_j]=\delta_{ij}$. The coupling $g$ gives rise to the
formation of polaritons (combined boson-qubit excitations) whose number
operator $\hat{N}=\sum_i (a^\dag_i a^\nag_i + \sigma^+_i\sigma^-_i)$ commutes
with $\hat{H}$. Hence, within this theoretical framework, the polariton
number is conserved and can be controlled via the chemical potential $\mu$.

The second term in Eq.~(\ref{eq:ham}) describes photon hopping between site
$i$ and site $j$ of the cavity array with amplitude $t_{ij}$.  At this point, the
hopping is completely general, and may involve any pair of lattice sites. In
previous work, only nn transfer along the directions of the lattice basis
vectors was considered, with $t_{ij}=-t$. We refer to this case as the {\it
  original} JCHM in the following.

Here we consider the JCHM with more complex hopping terms, motivated by
possible realizations using trapped ions in 1D Paul traps, and stripline
resonators on a 2D square lattice.

\subsection{Trapped ions}
The case of trapped ions implies a 1D model with frustrated
long-range hopping
\begin{equation}\label{eq:nnn}
  t_{ij} = t \frac{(-1)^{|i-j|}}{|i-j|^3} 
\end{equation}
arising from dipole-dipole interactions between confined ions
\cite{Ivanov1}. Here, $t$ denotes the nn hopping strength. The effects of
long-range hopping for trapped ions have been investigated using simple
mean-field like approximations \cite{Ivanov2}. However, the validity of
mean-field theory in such a 1D setting is not clear.  In particular, it fails
to predict the Kosterlitz-Thouless transition at the tip of the Mott lobe
which manifests itself as a non-analytic cusp in the phase boundary
\cite{PhysRevB.58.R14741,Ro.Fa.07}.  In this paper, we therefore address this
problem using numerical methods.

Due to the alternating sign in Eq.~(\ref{eq:nnn}), which gives rise to
frustration, QMC simulations are impracticable. Instead, we use the VCA, a
quantum cluster method which can be applied to frustrated systems and also to
compute single-particle excitation spectra. To benchmark the VCA for the case
of long-range hopping, we also present results for the case
\begin{equation}\label{eq:nnnabs}
  t_{ij} = \frac{t}{|i-j|^3}\,,
\end{equation}
\ie, without the alternating sign. Hamiltonian~(\ref{eq:ham}) with the
hopping integrals given by Eq.~(\ref{eq:nnnabs}) can be studied using QMC
simulations to obtain the exact phase boundary for the Mott-superfluid
transition.

\subsection{Stripline resonators}

The 2D JCHM may be realized in experiment using stripline resonators on a
square lattice. A simple configuration is shown in
Fig.~\ref{fig:lattice}. The hopping integral $t_{ij}=-t$ when
$i$ and $j$ are nn's along a lattice bond (as in previous
studies of the 2D JCHM), and $t_{ij}=-t'$ for diagonal nnn's
on every other plaquette. For any other pair of lattice sites we have $t_{ij}=0$.

In this circuit QED setup, the hopping rates $t$ and $t'$ are proportional
to the mutual capacitances of the resonators in Fig.~\ref{fig:lattice} \cite{PhysRevA.82.043811,NuKo11}.
By choosing a proper geometry and/or distance between pairs of resonators
on the diagonal of the array (determining $t'$) and pairs of resonators
on the lattice axes (determining $t$), the diagonal hopping rate $t'$
can be made much smaller with respect to the nn hopping rate
$t$ and vice versa. Thus, the ratio $t'/t$ can be engineered almost arbitrarily \cite{Ga.Ha.07}.
Here we consider the ratio $t'/t=1/2$. 

The freedom to tune the ratio $t'/t$ to
substantial values in circuit QED systems is in strong contrast to ultra-cold
atoms in optical lattices. For the latter, hopping integrals beyond nn pairs
are orders of magnitude smaller \cite{PhysRevLett.81.3108}. Moreover,
diagonal hopping is absent on hypercubic lattices because of the
orthogonality of the Wannier states. The small size of hopping integrals
beyond the nn terms makes their effect very small, with correspondingly few
studies available \cite{Ga.Ha.07,Za.Ko.10}.

The circuit QED model with hopping integrals $t$ and $t'$ as depicted in
Fig.~\ref{fig:lattice} can be simulated using the QMC method. For comparison,
we also present results for a 2D model with nn hopping $t$, and nnn hopping
$t''$ between sites separated by two lattice constants, along the $(1,0)$ and
$(0,1)$ directions.

\section{Methods}\label{sec:methods}

This section describes the methods used in this work in some detail.  In two
dimensions, we use exact QMC simulations, the VCA, and an analytical
linked-cluster approximation around the mean-field limit in two dimensions. Since
mean-field theory is not a valid starting point for an expansion in one
dimension, we show only numerical results (QMC, VCA) for that case.

\subsection{Quantum Monte Carlo}

In the absence of frustration, bosonic Hamiltonians of the
form of~(\ref{eq:ham}) can be studied by exact QMC simulations. A particularly
popular representation is the continuous-time stochastic series expansion
\cite{SandvikSSE1}; it has been applied to the JCHM model before
\cite{Zh.Sa.Ue.08,Pi.Ev.Ho.09,HoAiScPo11}.  We use the ALPS 1.3
implementation \cite{ALPS_I} of the stochastic series expansion with directed
loop updates \cite{SySa02,ALPS_DIRLOOP,PhysRevE.70.056705}. A maximum site occupation of six
photons is sufficient to make the truncation error negligible. The long-range
hopping defined by Eq.~(\ref{eq:nnnabs}) is treated by taking into account
all transfer processes with $|i-j|\leq L/2-1$ for system size $L$ (due to
periodic boundary conditions).

The Mott-superfluid phase boundary can be determined by calculating the
superfluid density 
\begin{equation}
  \rho_\text{s} = \frac{\las w^2\ras}{\beta D L^{D-2}}\,,
\end{equation}
where $D$ denotes the dimension of the lattice, $w$ is the winding number
\cite{PhysRevB.36.8343}, and $\beta$ is the inverse temperature. Exploiting
the scaling form of $\rho_s$ \cite{PhysRevB.40.546}, the critical point can
be obtained from simulations with different system sizes at inverse
temperature $\beta/L^z=\text{const.}$ The dynamical critical exponent of the
JCHM is $z=1$ for the fixed-density transition, and $z=2$ for the generic
transition \cite{HoAiScPo11}. For large enough values of the system size $L$,
curves for different $L$ values intersect at a single point
\cite{Zh.Sa.Ue.08,HoAiScPo11} which defines $t_\text{c}$ (for horizontal
scans in the $t,\mu$ phase diagram) respectively $\mu_\text{c}$ (for vertical
scans). We use system sizes up to $L=64$ in one dimension, and up to
$40\times40$ for the 2D models. The inverse temperatures were typically
$\beta/L^2=1/4$ for $z=2$, and $\beta/L=2$ or 4 for $z=1$. The resulting
accuracy for the phase boundary in units of $g$ is estimated to be better
than 0.0005 (below the symbol size used in the figures).

\subsection{Variational cluster approach}

The minus-sign problem resulting from the frustration induced by
choice~(\ref{eq:nnnabs}) for the hopping integrals motivates the use of an
alternative numerical method. The VCA is a cluster method; hopping
within a finite reference cluster is treated exactly, and hopping beyond the
cluster is taken into account perturbatively. For the relation to other
popular cluster methods such as the dynamical mean field theory, see \cite{PoAiDa03}.
The VCA \cite{PoAiDa03} can be applied for any
choice of hopping integrals in one and two dimensions and, also, permits 
calculation of excitation spectra.  It has been applied to the original JCHM with $t'=0$ in
\cite{Ai.Ho.Ta.Li.08} and \cite{PhysRevB.81.104303}, where in the latter case
the problem was mapped to an effective model. In one dimension, the VCA
yields quantitatively reliable results \cite{Ai.Ho.Ta.Li.08,PhysRevB.81.104303}. 
In 2D, accurate results can be obtained except for a small region around the
Mott lobe tip in the phase diagram. 

The quality of the approximation depends on the size of the cluster compared
to the correlation lengths of the problem, and the number of variational
parameters. Since larger clusters and more parameters have comparable impact,
we use only the boson energy in the cluster
as a parameter, and vary the cluster size. This is also motivated by the
presence of long-range hopping terms in the models considered here. For
more details see \cite{Ai.Ho.Ta.Li.08}.

Here we investigate Hamiltonian~(\ref{eq:ham}) using the VCA in its
formulation for bosonic systems \cite{KoDu06}. We work exclusively at zero
temperature. To deal with the long-range hopping defined by
Eqs.~(\ref{eq:nnn}) and~(\ref{eq:nnnabs}), we allow for hopping processes up
to a distance $L-1$, where $L$ is the cluster size used in the calculation.

On the technical side, we note that care is required when carrying out the
sums over wave vectors in the calculation of the grand potential
\cite{PoAiDa03,Ai.Ho.Ta.Li.08}. Far from the lobe tip, the sums converge
rapidly, whereas close to the tip an increasingly finer mesh is required. We
attribute this effect to the momentum dependence of the self-energy and
excitations. Deep in the Mott phase, the particle and hole bands are almost
completely flat, whereas close to the lobe tip they acquire a substantial
dispersion \cite{Ai.Ho.Ta.Li.08}. Given convergence, the VCA
yields the correct form of the phase boundaries, leading to improved results
for the 2D case compared to exact QMC data
\cite{Ai.Ho.Ta.Li.08,Zh.Sa.Ue.08}.

\subsection{Linked-cluster expansion}

We analytically calculate the photonic Matsubara Green's function
$G_{ij}(\tau;\tau')= - \langle {\mathcal{T}} a_i(\tau)
\bar{a}_j(\tau')\rangle$ with the time-ordering operator {${\cal{T}}$ and
  $\bar{a}_j(\tau')=e^{H\tau'}a_j^\dagger e^{-H\tau'}$ using a linked-cluster
  expansion in terms of local cumulants originally developed by Metzner
  \cite{Me91} for the Fermi-Hubbard model, and recently applied to the
  Bose-Hubbard model \cite{Oh.Pe.08} and the JCHM \cite{Sc.Bl.10}. After a
  Fourier transformation and analytic continuation, the inverse Green's
  function directly yields the excitation spectrum $\omega({\bf k})$
  via $G^{-1}({\mathbf k},\omega)=0$ and the phase boundary $t_c(\mu)$ via
  $G^{-1}({\mathbf 0},0)|_{t_c(\mu)}=0$. Key results are summarized
  below.

\subsubsection{Random phase approximation}

Within the mean-field random phase approximation (RPA), the full Green's
function is given by \cite{Sc.Bl.09}
\begin{eqnarray}\label{eq:rpa}
  G({\bf k},\omega)=\frac{G^0(\omega)}{1-J({\bf k}) G^0(\omega)}
\end{eqnarray}
with the local one-particle cumulant
\begin{equation}\label{eq:cum1}
  G^0(\omega)=\sum_{\sigma=\pm} \frac{z_{n+1}^{-,\sigma}}
  {\Delta_{n+1}^{-,\sigma} - \omega}-\frac{z_{n}^{\sigma,-}}{\Delta_{n}^{\sigma,-} - \omega}\,.
\end{equation}
The hole/particle spectral weights and bare excitation energies in the
numerator and denominator in Eq.~(\ref{eq:cum1}) are given in the
Appendix.  In Eq.~(\ref{eq:rpa}), $J({\bf k})$ denotes the bare
dispersion. For the 2D circuit QED model in Fig.~\ref{fig:lattice},
\begin{equation}
  J({\bf k})=2 t (\cos{k_x} + \cos{k_y}) + 2 t' \cos{k_x k_y}\,.
\end{equation}

From Eq.~(\ref{eq:rpa}), we can derive analytical expressions for the phase
boundary and the excitation spectrum. For example, the RPA phase boundary for
the $t,t'$ model at a fixed ratio $R=t'/t$ is given by $1/t_c=4G^0(0)
(1+R/2)$. Thus, on the mean-field level, nnn hopping simply rescales the nn
hopping according to $t_c\mapsto t_c/(1+R/2)$.  Hence, a positive (negative)
nnn hopping decreases (increases) the size of the Mott lobes and makes
mean-field theory a better (worse) approximation. The RPA is equivalent
to the cluster perturbation theory (VCA without variational parameters) for
a single-site cluster \cite{KoDu06}.

\subsubsection{One-loop approximation}

The mean-field RPA corresponds to a summation of all self-avoiding walks
through the lattice.  In order to take into account the leading quantum
correction one can additionally include all one-time forward/backward hopping
processes between two neighboring sites.  The resulting so-called {\it one-loop
  approximation} was previously found to be in excellent agreement with
numerical results for the phase diagram for both the Bose-Hubbard model
\cite{Oh.Pe.08} and the original JCHM~(\ref{eq:ham}) with $t'=0$ \cite{Sc.Bl.09}.

\section{Results}\label{sec:results}

To calculate the phase diagram of the models discussed in
Sec.~\ref{sec:model}, we vary the nn hopping strength $t$ common to all
models, and scale any additional hopping integrals accordingly. In the
following, we use $g$ as the unit of energy, and consider
$\omega_\text{b}=\omega_\text{q}=1$ (resonance condition). We also set
$\hbar$, $k_\text{B}$ and the lattice constant to 1.

\subsection{Mott-superfluid transition}

Similarly to cold atoms in optical lattices, the JCHM can be tuned across the
Mott-superfluid transition by changing the ratio $t/g$. We therefore expect
the additional hopping terms discussed in Sec.~\ref{sec:model} to modify the
extent of the Mott insulating region in the phase diagram. Because the Mott
phase is of particular interest as an initial state for possible applications
in quantum computing and quantum information, we determine the size of the
first (largest) Mott lobe with polariton density $n=1$.

\begin{figure}[t]
  \centering
  \includegraphics[width=0.45\textwidth]{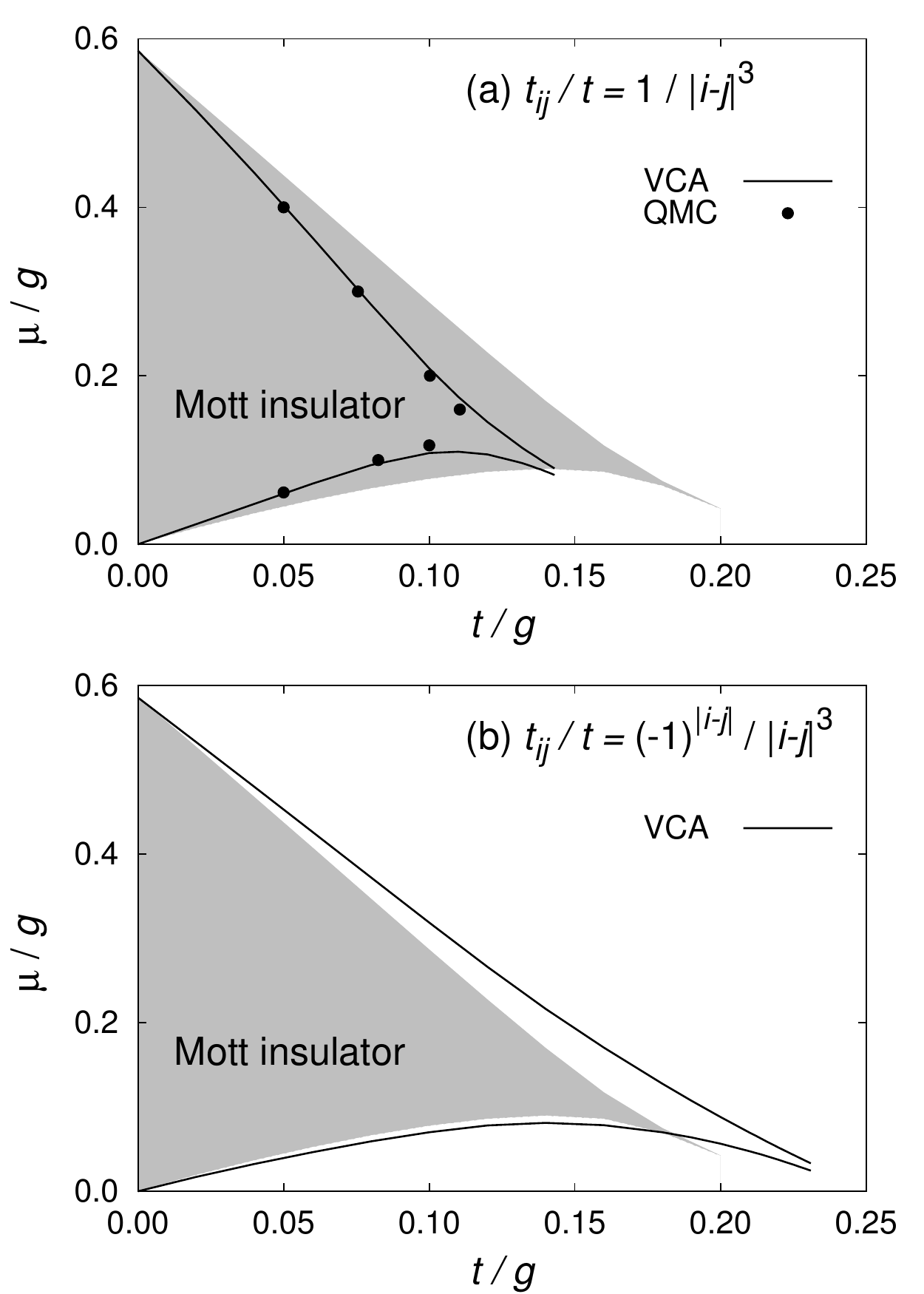}
  \caption{\label{fig:1d_qmc}%
    (Color online)
    Zero-temperature Mott lobe with density $n=1$ of the 1D JCHM, showing the
    effect of (a) long-range hopping defined by Eq.~(\ref{eq:nnnabs}), and
    (b) sign-alternating long-range hopping defined by Eq.~(\ref{eq:nnn}), as
    appropriate for an experimental realization based on trapped ions.  Case
    (a) can be investigated by QMC simulations (symbols), and the phase
    boundary is compared to the VCA (lines, for a cluster size $L=8$). For
    reference, we also show the exact result for the original JCHM with
    nearest-neighbor hopping only \cite{Ro.Fa.07} (shaded regions). Case
    (b) is not accessible for QMC, and we rely on the VCA for the phase
    diagram.}
\end{figure}

\subsubsection{One dimension}

We begin with the 1D model with long-range hopping given by
Eq.~(\ref{eq:nnnabs}), which represents an interesting theoretical problem
for two reasons. First, it allows us to benchmark the VCA against exact QMC
results for the novel case of long-range hopping. Second, the long-range
hopping is expected to mimic the effect of increasing the lattice
coordination $Z$, while keeping the lattice topology and dimension
unchanged. Previous VCA calculations for the JCHM have demonstrated a
crossover from Kosterlitz-Thouless behavior (reflected in a strongly
non-parabolic shape of the phase boundary, including reentrant behavior
\cite{PhysRevB.58.R14741}) on a 1D chain, to a mean-field like, parabolic
phase boundary for the 2D square lattice \cite{Ai.Ho.Ta.Li.08}. This
crossover is also visible upon comparing Figs.~\ref{fig:1d_qmc}
and~\ref{fig:2d_qmc}. Mean-field behavior in a 1D matter-light system is also
suggested by results for a two-component Bose-Hubbard model coupled to a
global photon mode (as in the Dicke model) \cite{BhHoSiSi08}; the coupling
gives rise to photon-mediated long-range interaction.

Figure~\ref{fig:1d_qmc}(a) shows the Mott-superfluid phase boundary for the
first Mott lobe. The long-range hopping substantially reduces the
extent of the insulating phase, as compared to the original JCHM (shaded
regions, results taken from \cite{Ro.Fa.07}). Due to the particular shape of
the phase boundary in one dimension, the reduction of the critical hopping due to
$t_{ij}$ can be very large, about 50 percent for $\mu/g=0.1$.

The Kosterlitz-Thouless phase transition at a fixed polariton density, which
occurs at the tip of the Mott lobe, is difficult to study by the
QMC method. Away from the lobe tip, the critical points from QMC agree
well with the VCA results. The approximate cluster approach underestimates
the effect of quantum fluctuations for larger $t/g$, %(near the lobe tip),
leading to a slightly larger Mott phase than found by QMC as visible in
Fig.~\ref{fig:1d_qmc}(a) for the QMC data points located at $t/g\gtrsim
0.1$. This finding is in accordance with the cluster size dependence of the
VCA Mott lobe in \cite{Ai.Ho.Ta.Li.08}, where satisfactory convergence was
achieved for $L=8$ (the same size as used here). Far away from the lobe tip,
for $t/g<0.1$ in Fig.~\ref{fig:1d_qmc}(a), the transition is essentially
driven by density fluctuations, and the VCA and QMC results agree almost
perfectly.

\begin{figure}[t]
  \centering
  \includegraphics[width=0.45\textwidth]{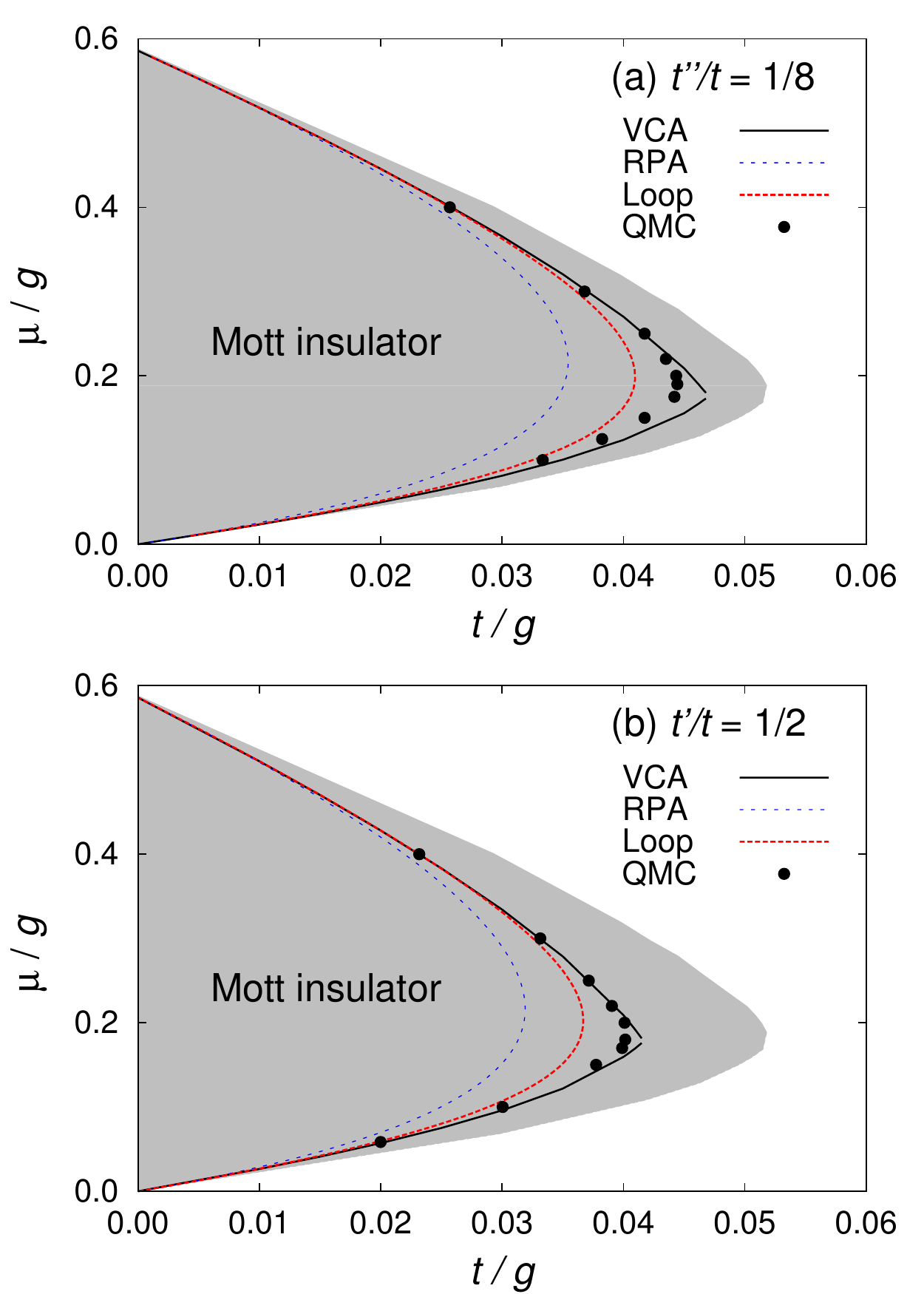}
  \caption{\label{fig:2d_qmc}%
    (Color online) Zero-temperature Mott lobe with density $n=1$ of the 2D
    JCHM, showing the effect of (a) next-nearest-neighbor hopping $t''=t/8$
    (see Sect.~\ref{sec:model}) and (b) {\it diagonal} next-nearest-neighbor
    hopping $t'=t/2$ as appropriate for an experimental realization based on
    stripline resonators (see Fig.~\ref{fig:lattice} and
    Sec.~\ref{sec:model}). We show results from QMC simulations (symbols),
    the VCA (lines, $L=8$), the mean-field theory (RPA), and the one-loop
    approximation. For reference, we also include the $t'=t'=0$ QMC results
    for the original 2D JCHM model (shaded regions, taken from
    \cite{Zh.Sa.Ue.08}). }
\end{figure}

Concerning a possible change of the universality class in the presence of
long-range hopping, neither the QMC nor the VCA results provide any evidence
of such a scenario. The shape of the Mott lobe for the model with long-range
hopping is reminiscent of the phase boundary of the original JCHM
(shaded region). Besides, the QMC results for the superfluid density
(not shown) exhibit scaling with dynamical critical exponent $z=1$. The
absence of mean-field behavior despite long-range hopping is a consequence
of the fast decay with distance, $t_{ij}\sim |i-j|^{-3}$, see
Eq.~(\ref{eq:nnnabs}).

Turning to the model for trapped ions, with sign-alternating long-range
hopping as given by Eq.~(\ref{eq:nnn}), we note that the QMC method suffers
from a severe minus-sign problem. Relying on the accuracy of the VCA
established in Fig.~\ref{fig:1d_qmc}(a) and in previous work, we show the VCA
phase boundary in Fig.~\ref{fig:1d_qmc}(b). The sign-alternating hopping has
an effect opposite to the choice~(\ref{eq:nnnabs}). The extent of the Mott is
noticeably increased compared to the model with nn hopping only.  The largest
change of the critical hopping again occurs for $\mu/g\approx0.1$, with an
increase of about 50 percent. The qualitative shape of the phase boundary is
unchanged.

\subsubsection{Two dimensions}

We now turn to the results for the 2D models discussed in
Sec.~\ref{sec:model}. Since there is no frustration in either case, we can
study the Mott-superfluid transition exactly using the QMC method. The
findings are compared to the VCA, and to analytical approximations.

Figure~\ref{fig:2d_qmc}(a) contains the results for the $t,t''$ model, with
$t''/t=1/8$ [motivated by taking a $1/r^3$ dependence as in
Eq.~(\ref{eq:nnnabs}) but keeping only the first two terms]. The hopping
$t''$ causes a 10 percent reduction of the critical hopping close to the tip
as compared to the model with nn hopping only \cite{Zh.Sa.Ue.08}.  The VCA
provides a remarkably good description of the whole phase boundary.  In particular, the
spurious inflection points visible in earlier work
\cite{Ai.Ho.Ta.Li.08,Zh.Sa.Ue.08} are absent. As in one dimension, the
underestimation of spatial fluctuations leads to slightly larger values of
the critical hopping for the Mott-superfluid transition. Conversely, the RPA
mean-field result severely underestimates the effect of fluctuations, a
deficiency which is to a large extent remedied within the one-loop
approximation.

Results for the 2D circuit QED model with hoppings $t'/t=1/2$ are shown in
Fig.~\ref{fig:2d_qmc}(b). The effect of $t'$ is qualitatively very similar to
that in Fig.~\ref{fig:2d_qmc}(a), with the larger value of $t'$ compared to $t''$
leading to a stronger decrease of the critical hopping close to the tip
(about 20 percent). The agreement between the different methods, in
particular the VCA and QMC, is even better than for the $t,t''$ model. Hence,
except for details such as critical exponents \cite{HoAiScPo11}, both the VCA
and the one-loop approximation may be used for semi-quantitative analyses of
the 2D JCHM at substantially smaller computational cost than for the QMC.

\subsection{Excitation spectra}

\begin{figure}
  \centering
  \includegraphics[width=0.5\textwidth,clip]{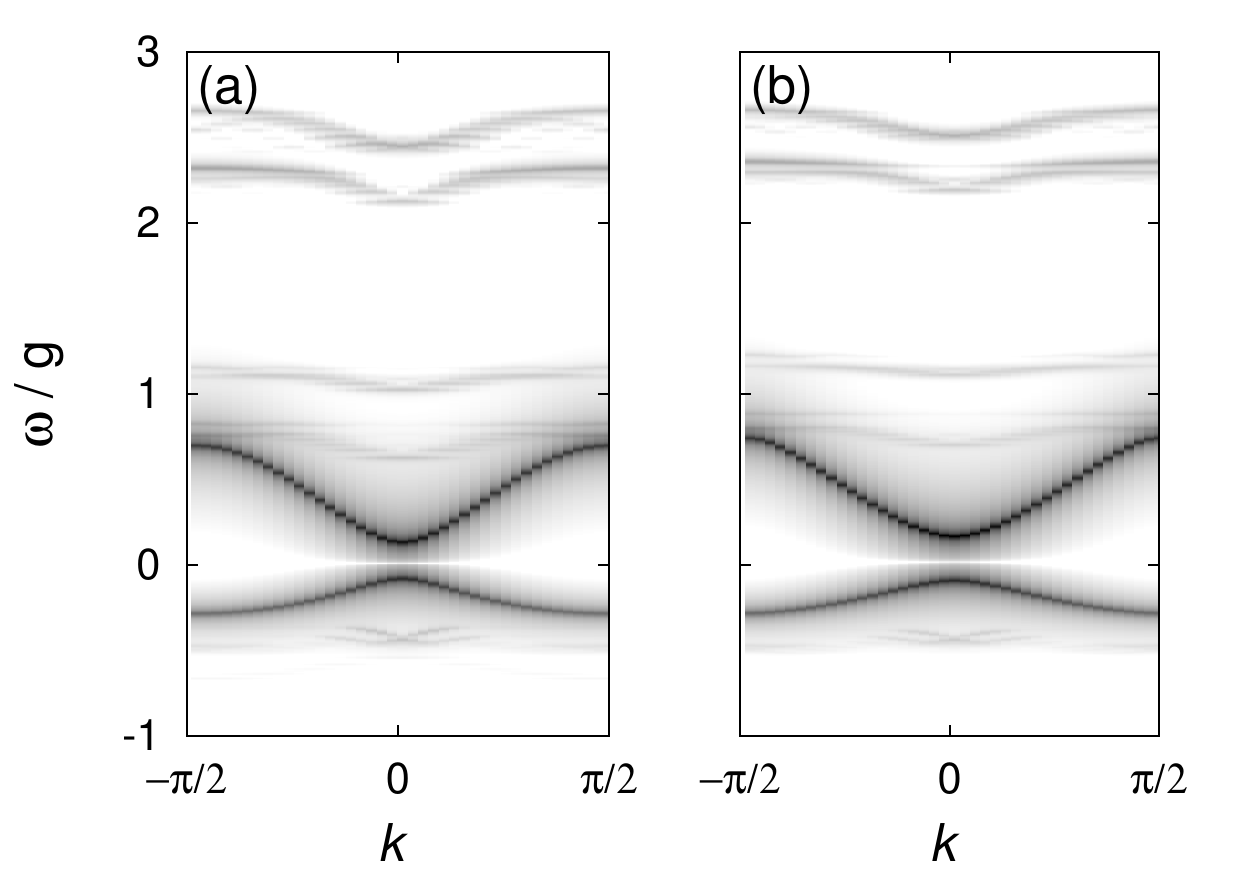}
  \caption{\label{fig:spec_1D}%
    (Color online)
    Zero-temperature excitation spectra of the 1D JCHM at zero detuning with
    (a) nearest-neighbor hopping $t$, and (b) sign-alternating long-range
    hopping $t_{ij}$, relevant for trapped ions [Eq.~(\ref{eq:nnn})]. Results
    are from the VCA with cluster size $L=8$, and using $t/g=0.1$,
    $\mu/g=0.15$, see Fig.~\ref{fig:1d_qmc}(b).}
\end{figure}

\begin{figure}[t]
  \centering
  \includegraphics[width=0.5\textwidth,clip]{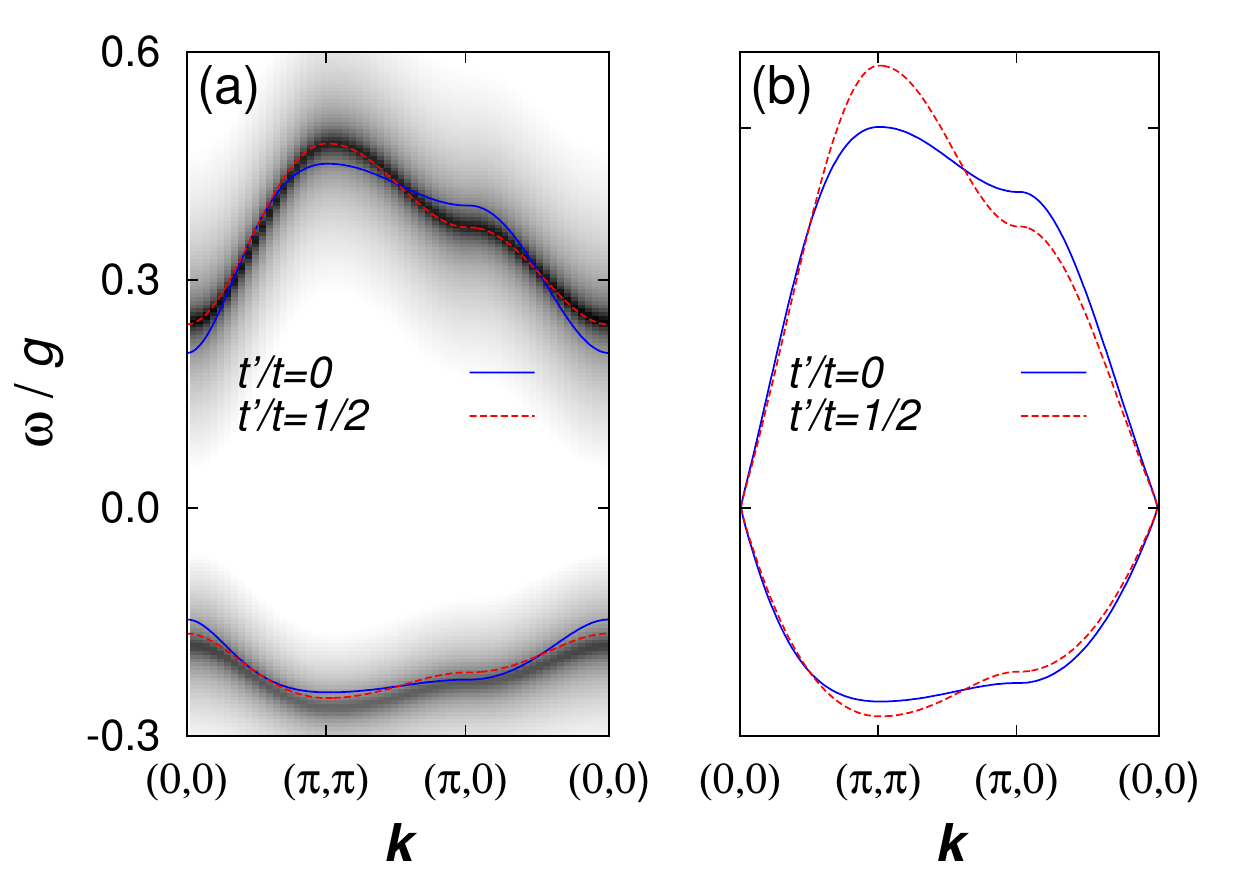}
  \caption{\label{fig:spec_2D}%
    (Color online) Zero-temperature excitation spectra for the 2D JCHM at
    zero detuning with $t'=0$ (solid lines), or $t'/t=1/2$ [dashed lines and
    intensity plot in (a)] as appropriate for a circuit-QED setup (see
    Fig.~\ref{fig:lattice} and Sec.~\ref{sec:model}).  (a) Parameters deep in
    the Mott lobe, using $t/g=0.02$, $\mu/g=0.22$. (b) Parameters exactly at
    the Mott lobe tip, using $t/g=0.04$ (for
    $t'=0$) respectively $t/g=0.032$ (for $t'/t=1/2$), and $\mu/g=0.22$, see
    Fig.~\ref{fig:2d_qmc}(b).  Lines are RPA results for the particle/hole
    dispersion. The density plot in (a) shows VCA results based on a
    $2\times2$ cluster.  }
\end{figure}

Single-particle excitation spectra are directly accessible in cavity QED via
photoemission spectroscopy. As demonstrated previously
\cite{Ai.Ho.Ta.Li.08,Pi.Ev.Ho.09,Sc.Bl.10}, they allow us to identify the system
state via the presence or absence of a Mott gap and to calculate
effective particle and hole masses. We therefore study here the effect of the
additional hopping processes in models for trapped-ion and circuit-QED
realizations of the JCHM.

Figure~\ref{fig:spec_1D} compares the excitation spectrum deep in the Mott
phase for the cases of the original 1D JCHM, Fig.~\ref{fig:spec_1D}(a), and
the model with sign-alternating long-range hopping,
Fig.~\ref{fig:spec_1D}(b). Results were obtained using the VCA.
The four gapped branches of the original model persist in the presence of
long-range hopping; the two low-lying branches correspond to conventional
particle/hole excitations \cite{Ai.Ho.Ta.Li.08}, whereas the high-energy
branches are so-called conversion modes, which arise from the composite
nature of polaritons \cite{Sc.Bl.09}. The upper modes have a very small
bandwidth and low spectral weight. Comparing Figs.~\ref{fig:spec_1D}(a) and
(b), we conclude that the effect of the long-range hopping is negligible.  A
comparison close to the lobe tip is difficult in one dimension, because of
the different extent of the Mott phases, see Fig.~\ref{fig:1d_qmc}(b).

Results for the excitation spectra of the 2D circuit-QED model are shown in
Fig.~\ref{fig:spec_2D}, focusing on low-energy modes. Deep in the Mott
phase, we find perfect agreement between the VCA and the RPA. The
renormalization of the Mott-superfluid transition (including the Mott gap)
due to $t'$ [see Fig.~\ref{fig:2d_qmc}(b)] is visible from the particle and
hole bands close to the $\bm{\Gamma}$ point.

Finally, we consider the spectrum exactly at the lobe tip. Because our
formulation of the VCA is restricted to the gapped Mott phase, we only show
the corresponding RPA results in Fig.~\ref{fig:spec_2D}(b). The ratio $t/g$
has been tuned to the lobe tip for both $t'=0$ and $t'>0$. The RPA dispersion
becomes relativistic with gapless, linear modes in both cases, and noticeable
differences due to $t'$ only far away form the $\Gamma$ point and hence at
high energies. In particular, the sound velocities are practically
identical. The presence of linear excitations at the lobe tip is a signature
of the fixed-density transition with dynamical critical exponent $z=1$.
The value $z=1$ has recently been demonstrated using QMC simulations
\cite{CaSa.GuSo.Pr.Sv.07,HoAiScPo11} for the case $t''=t'=0$ (original
JCHM). As expected, the universality does not change in the presence of additional hopping terms.

\section{Summary}\label{sec:summary}

We have investigated the effects of additional hopping terms beyond the
generic nearest-neighbor transfer in the Jaynes-Cummings-Hubbard model in one
and two dimensions. Such hopping terms arise when considering possible
experimental realizations based on trapped ions in a linear Paul trap, or
stripline resonators on a square lattice.
Using numerical and analytical methods, we have shown that the phase diagram
is modified substantially. Compared to the original model with
nearest-neighbor hopping only, the Mott lobe with density 1 becomes enlarged
in the case of trapped ions, but is reduced for the case of stripline
resonators. This effect is particularly pronounced at selected chemical
potentials in one dimension, due to the shape of the Mott lobe.  In contrast,
excitation spectra are only very weakly affected by the additional hopping
terms, especially at low energies. In particular, the sound velocity at the
lobe tip as well as the universality class of the JCHM remain unchanged.

An interesting problem arises from the possibility of tuning the hopping ratio
$t'/t$ in a circuit-QED setup. For $t'\gg t$, the lattice separates into
independent, diagonal, one-dimensional chains. By varying $t'/t$, one could
thus observe a crossover from one to two dimensions, \ie, from
Kosterlitz-Thouless behavior to mean-field like behavior.

\begin{acknowledgments}
  We are grateful F.~Assaad, G.~Blatter, A.~Houck, J.~Keeling, P.~Pippan and
  M.~Troyer for useful discussions. This work made use of the ALPS
  applications \cite{ALPS_DIRLOOP}. MH was supported by DFG FG1162, and
  acknowledges the hospitality of ETH Zurich. LP was supported by the SNSF
  under Grant No. PZ00P2-131892/1.
\end{acknowledgments}

\vspace*{1em}
\appendix*\section{Atomic-limit results}\label{sec:app}

In the atomic limit ($t_{ij}=0$) the eigenstates of the Hamiltonian~(\ref{eq:ham})
are the dressed polariton states labeled by the polariton number $n$ and
upper/lower branch index $\sigma=\pm$.  For $n>0$ they can be written as a
superposition of a Fock state with $n$ photons plus atomic ground state $|n,
\text{g}\rangle$ and $(n-1)$ photons with the atom in its excited state
$|(n-1), \text{e}\rangle$,
\begin{eqnarray}
|n +\rangle &=& \sin\theta_n |n\,,\text{g}\rangle + \cos\theta_n |(n-1)\,,\text{e}\rangle\,, \nonumber\\
|n -\rangle &=& \cos\theta_n |n\,,\text{g}\rangle -  \sin\theta_n |(n-1)\,,\text{e}\rangle\,,
\end{eqnarray}
where
\begin{equation}
\tan \theta_n=2 g \sqrt{n}/(\delta+2\chi_n)
\end{equation}
with
\begin{equation}
\chi_n=\sqrt{g^2 n + \delta^2/4}
\end{equation}
and the detuning $\delta=\omega_\text{b}-\omega_\text{q}$. 
The eigenvalues are
\begin{equation}
\epsilon_n^{\sigma}=-(\mu-\omega_\text{b}) n - \delta/2 +\sigma\,\chi_n\,,\quad \sigma=\pm\,.
\end{equation}
%
%\vspace*{1em}\\
The zero polariton state $|0-\rangle=|0\,,g\rangle$ is a special case with $\epsilon_0^-=0$. 
Upper and lower polariton energies are separated by the Rabi splitting $\Omega_n=2\chi_n$. 
The spectral weights in Eq.~(\ref{eq:cum1}) are then defined as
\begin{equation}
  z^{\sigma,\sigma'}_n=(f_{n}^{\sigma,\sigma'})^2
\end{equation}
with the matrix elements $f_{n}^{\sigma,\sigma'}=\langle n\sigma|a^\dagger| (n-1) \sigma'\rangle$. The bare excitation energies are given by
\begin{equation}
  \Delta^{\sigma,\sigma'}_n=\epsilon_{n}^\sigma-\epsilon_{n-1}^{\sigma'}\,.
\end{equation}
%

%\bibliographystyle{../prsty} 
%\bibliography{../bibliography}

\end{document}